\documentclass[reprint,superscriptaddress,amsmath,amssymb,aps,prl,nofootinbib]{revtex4-2}
\pdfoutput=1
\usepackage{graphicx}
\usepackage{bm}
\usepackage{amsmath}
\usepackage{hyperref}
\newtheorem{definition}{Definition}

\begin{document}

\title{Acoustic Firewalls: Analogue Gravity Perspective on the AMPS Paradox}

\author{N.~S. Akintsov}
\email{akintsov777@ntu.edu.cn}
\affiliation{School of Artificial Intelligence and Computer Science,
 Nantong University, Nantong 226019, China}

\author{A.~P. Nevecheria}
\email{artiom.nevecherya@gmail.com}
\affiliation{Department of Mathematical and Computer Methods,
 Kuban State University, Krasnodar 350040, Russia}

\author{S.~N. Andreev}
\email{andreev@cir-innovations.ru}
\affiliation{Joint-Stock Company ``Center for Research and Development'',
 Moscow 101000, Russia}

\author{Qing-Hua Qin}
\email{qinghua.qin@smbu.edu.cn}
\affiliation{Institute of Advanced Interdisciplinary Technology,
 Shenzhen MSU-BIT University, Shenzhen 518172, China}

\date{\today}

\begin{abstract}
The monogamy of quantum entanglement, applied by Almheiri--Marolf--Polchinski--Sully
(AMPS) to black holes, obstructs a smooth horizon vacuum after the Page time. We transcribe
this argument to Hawking-like phonon radiation from a sonic horizon in the Unruh acoustic
metric. An exact purity identity shows that post-Page-time unitarity forces the entanglement
between an outgoing phonon and its interior partner to vanish, selecting a non-Hadamard
(Boulware-like) phonon state---which we define as an \emph{acoustic firewall}. Its renormalized
stress tensor differs from the smooth state by a constant, negative near-horizon flux, and the
thermal-atmosphere energy density it removes---measured by a static calorimeter---grows as
$(\delta r)^{-2}$ in the radial coordinate toward the horizon (singular in the free-fall frame), cut
off at the healing length. The construction is
kinematic and does not resolve the information paradox; it yields one concrete, falsifiable
prediction: a differential phonon-calorimetry signal
$\mathcal{R}(\delta r)=|\Delta\mathcal{E}|/\mathcal{E}^{(0)}\!\to(\ell_\kappa/\delta r)^{2}$,
present only after the analogue Page time in a Bose--Einstein condensate.
\end{abstract}

\maketitle

\emph{Introduction.}---Hawking's prediction that black holes radiate thermally~\cite{Hawking1975}
places general relativity and quantum mechanics in tension: unitary evolution of the emitted
radiation appears incompatible with a smooth horizon and local effective field theory. Almheiri,
Marolf, Polchinski, and Sully (AMPS) sharpened this into the \emph{firewall} argument~\cite{AMPS2013}:
the \emph{monogamy of entanglement}~\cite{CKW2000} makes three individually reasonable postulates---
unitarity, semiclassical local physics at the horizon, and no drama for an infalling observer---
mutually inconsistent after the Page time~\cite{Page1993}. The modern island/entanglement-wedge
program recovers a unitary Page curve~\cite{AEMM2019,Penington2020,AlmheiriRMP2021}, but the
physical status of the horizon remains debated~\cite{MaldacenaSusskind2013,Mathur2009}.

Analogue gravity~\cite{BLV2005,Volovik2003} offers laboratory horizons: Unruh showed that sound in a moving
fluid obeys a curved-space wave equation with an acoustic (``sonic'') horizon where the flow turns
supersonic~\cite{Unruh1981}. Sonic horizons in Bose--Einstein condensates (BECs) emit thermal
phonons~\cite{Garay2000,Carusotto2008,Recati2009}, and this analogue Hawking radiation---together
with its entanglement---has been observed~\cite{Steinhauer2016,deNova2019,Kolobov2021} and in
classical surface-wave systems~\cite{Weinfurtner2011}. These experiments realize the \emph{kinematic}
ingredients of the AMPS setup: a thermal spectrum, an entangled partner structure, and a monotonic
entropy growth. It is therefore natural to ask whether the monogamy obstruction that underlies the
firewall has an analogue at a sonic horizon, and whether it leaves a measurable signature. We answer
both affirmatively and identify a falsifiable observable. We stress at the outset that the analogy is
purely kinematic: it involves no Einstein equations and does not claim to resolve the information paradox.

\begin{figure}[t]
 \includegraphics[width=\columnwidth]{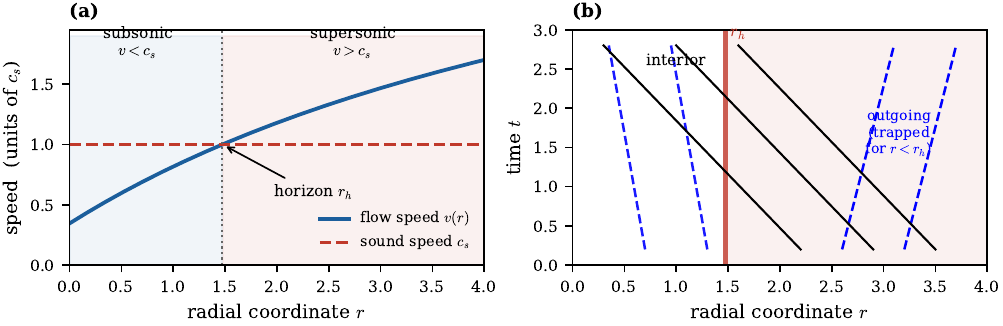}
 \caption{\label{fig:horizon}Sonic horizon. (a)~Flow speed $v(r)$ crosses the sound speed $c_s$ at
 the horizon $r_h$, separating subsonic ($v<c_s$) and supersonic ($v>c_s$) regions. (b)~Ray diagram:
 ingoing null rays (solid) cross $r_h$; outgoing rays (dashed) are trapped for $r<r_h$, the acoustic
 analogue of the interior.}
\end{figure}

\emph{Acoustic black-hole model.}---A barotropic, irrotational flow of density $\rho$ and velocity
$\bm{v}$ carries linearized phonons $\hat\phi$ that obey a massless Klein--Gordon equation in the
acoustic metric~\cite{Unruh1981,BLV2005}
\begin{equation}
 ds^2=\frac{\rho}{c_s}\!\left[-(c_s^2-v^2)\,dt^2-2v\,dr\,dt+dr^2+r^2 d\Omega^2\right],
 \label{eq:metric}
\end{equation}
with $c_s$ the local sound speed. Equation~\eqref{eq:metric} shares the null structure of the
Painlev\'e--Gullstrand form of Schwarzschild spacetime under $v(r)\leftrightarrow\sqrt{2GM/r}$; the
sonic horizon sits at $v(r_h)=c_s$ [Fig.~\ref{fig:horizon}]. Its surface gravity and analogue
Hawking temperature are
\begin{equation}
 \kappa=\frac{1}{2c_s}\,\partial_r\!\left(c_s^2-v^2\right)\big|_{r_h},\qquad
 T_{\rm ac}=\frac{\hbar\kappa}{2\pi k_B},
 \label{eq:Tac}
\end{equation}
In a BEC governed by the Gross--Pitaevskii equation, $c_s=\sqrt{g n_0/m}$ with $g=4\pi\hbar^2 a_s/m$, and
the healing length $\xi=\hbar/(\sqrt{2}\,m c_s)$ sets the short-distance (``trans-Planckian'') cutoff of the analogy.
The Bogoliubov map between in- and out-vacua produces a thermal phonon flux at $T_{\rm ac}$: each
outgoing mode $\hat a^{\rm out}_\omega$ is created together with an interior partner $\hat b_\omega$
in the two-mode squeezed state
\begin{equation}
 |\Psi\rangle=\sqrt{1-q}\,\sum_{n=0}^{\infty}q^{\,n/2}\,|n\rangle_{\rm out}|n\rangle_{\rm in},
 \qquad q=e^{-\hbar\omega/k_B T_{\rm ac}},
 \label{eq:tms}
\end{equation}
whose exterior reduced state is exactly thermal. This partner entanglement, verified experimentally
via density--density correlations~\cite{Balbinot2008,Steinhauer2016,deNova2019} and quantified
theoretically by the entanglement negativity of the emitted radiation~\cite{ChandranFischer2026}, is
the starting point for the acoustic monogamy argument.

\emph{Phonon monogamy and the acoustic firewall.}---Tracing out the interior in Eq.~\eqref{eq:tms}
gives a thermal exterior state whose von Neumann entropy grows monotonically as phonons are emitted---
the analogue of Bekenstein--Hawking entropy growth. Label the early phonon radiation (emitted before
the analogue Page time $t_{\rm Page}$) by $R$, a late outgoing mode by $A\equiv\hat a^{\rm out}_\omega$,
and its partner by $B\equiv\hat b_\omega$. Three conditions, each individually well motivated, cannot
hold together:
\begin{itemize}
 \item[(C1)] \emph{Unitarity.} The global phonon state is pure, so for $t>t_{\rm Page}$ mode $A$ must
 be nearly maximally entangled with $R$ to purify the growing entropy~\cite{Page1993}.
 \item[(C2)] \emph{Local field theory at the horizon.} The Bogoliubov (Hadamard) vacuum requires $A$
 maximally entangled with its partner $B$ for a regular phonon state at $r_h$~\cite{Recati2009}.
 \item[(C3)] \emph{Monogamy.} $A$ cannot be simultaneously maximally entangled with both $R$ and
 $B$~\cite{CKW2000,OsborneVerstraete2006}.
\end{itemize}
This is the acoustic transcription of the AMPS paradox. It is made exact by an identity, not an
inequality: for the pure tripartite state of $A$, $B$, and $R$, the mutual informations obey
\begin{equation}
 I(A{:}R)+I(A{:}B)= 2S(A),
 \label{eq:purity}
\end{equation}
which follows directly from $S(AR)=S(B)$ and $S(AB)=S(R)$. Post-Page-time unitarity requires
$I(A{:}R)\to 2S(A)$, whence
\begin{equation}
 I(A{:}B)\to 0 .
 \label{eq:disent}
\end{equation}
The partner entanglement that guarantees a Hadamard state at $r_h$ is thereby destroyed. We give the
resulting object a precise name.

\begin{definition}[Acoustic firewall]
An \emph{acoustic firewall} is a stationary, non-Hadamard state of the linearized phonon field in the
acoustic metric~\eqref{eq:metric} that is consistent with Eq.~\eqref{eq:disent}: the outgoing/interior
partner entanglement vanishes, $I(A{:}B)\to0$, so that its renormalized stress tensor
$\langle\hat T_{\mu\nu}\rangle$, while differing from the Hadamard state only by a finite (constant) null
flux, has a proper energy density that is singular in the free-fall frame at $r_h$.
\end{definition}

The term is used only in this technical sense---a phonon-sector state, not a gravitational firewall.
Equation~\eqref{eq:disent} does not select a unique state; the minimal representative consistent with it
is the Boulware-like state (no outgoing partner correlations across $r_h$), which we adopt throughout.
The far-field thermal energy-density scale of the partner structure is
\begin{equation}
 \mathcal{E}_{\rm fw}^{(0)}\sim\frac{(k_B T_{\rm ac})^4}{\hbar^3 c_s^3},
 \label{eq:E0}
\end{equation}
the Stefan--Boltzmann density of a phonon gas at $T_{\rm ac}$; for the $^{87}$Rb parameters below,
$\mathcal{E}_{\rm fw}^{(0)}\sim5\times10^{-20}~{\rm J/m^3}$.

\begin{figure}[t]
 \includegraphics[width=\columnwidth]{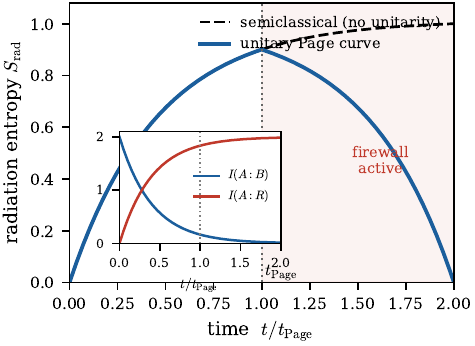}
 \caption{\label{fig:page}Analogue Page curve. Semiclassical entropy (dashed) grows monotonically;
 unitarity (solid) turns it over at $t_{\rm Page}$. Shaded: post-Page regime where the acoustic
 firewall is active. Inset: mutual informations $I(A{:}B)$ and $I(A{:}R)$ cross at $t_{\rm Page}$
 [Eqs.~\eqref{eq:purity}--\eqref{eq:disent}].}
\end{figure}

\emph{Near-horizon energy density.}---For the Hadamard (Hartle--Hawking--Unruh) state the point-split
renormalized $\langle\hat T_{uu}\rangle$ component equals the finite $(1+1)$-dimensional Hawking
luminosity~\cite{BLV2005},
\begin{equation}
 \langle\hat T_{uu}\rangle_{\rm H}=L_H=\frac{\pi k_B^2 T_{\rm ac}^2}{12\hbar}
 =\frac{\hbar\kappa^2}{48\pi},
 \label{eq:lumin}
\end{equation}
which we verify symbolically. The disentangled (Boulware-like) state of Eq.~\eqref{eq:disent} differs
from the Hadamard one by a constant, negative near-horizon flux $-\hbar\kappa^2/48\pi$, derived from first
principles by the trace-anomaly method (Supplemental Material); its renormalized components are finite in
static coordinates, but the proper (static-frame) energy density Tolman-enhances toward $r_h$ and
diverges in the free-fall frame---an observer dependence of the phonon content characteristic of BEC
analogue spacetimes~\cite{FedichevFischer2003,FedichevFischer2004}. Severing the partner entanglement thus \emph{removes} the near-horizon
thermal atmosphere: the Hadamard (pre-Page) atmosphere has proper density
$\mathcal{E}_{\rm pre}=(k_B T_{\rm loc})^4/(\hbar^3 c_s^3)$, $T_{\rm loc}=T_{\rm ac}c_s/\sqrt{c_s^2-v^2}$,
whereas the Boulware-like (post-Page) state is nearly empty ($\mathcal{E}_{\rm post}\!\approx\!0$).
Differential calorimetry across $t_{\rm Page}$ isolates the change
$\Delta\mathcal{E}=\mathcal{E}_{\rm post}-\mathcal{E}_{\rm pre}$, cancelling static backgrounds. With
$c_s^2-v^2\simeq2c_s\kappa\,\delta r$ ($\delta r=r-r_h$ the radial coordinate distance) near the horizon,
\begin{equation}
 |\Delta\mathcal{E}(\delta r)|=\mathcal{E}_{\rm fw}^{(0)}
 \left(\frac{c_s^2}{c_s^2-v^2}\right)^{\!2}
 \xrightarrow{\ \delta r\to0\ }\ \mathcal{E}_{\rm fw}^{(0)}\left(\frac{\ell_\kappa}{\delta r}\right)^2 ,
 \label{eq:diverge}
\end{equation}
a near-complete \emph{depletion} of the atmosphere ($\mathcal{E}_{\rm post}\!\approx\!0$, so
$|\Delta\mathcal{E}|\simeq\mathcal{E}_{\rm pre}$), set by the surface-gravity length
$\ell_\kappa=c_s/(2\kappa)$; the exponent $-2$ is a Stefan--Boltzmann extension of the rigorous $(1+1)$D
result ($-1$; Supplemental Material), fixed by the three-dimensionality of the phonon gas and the linear
vanishing of $g_{tt}$. The growth is formal---regulated at the healing length $\xi$ where the acoustic
metric fails~\cite{RibeiroBaakFischer2022}, so $\mathcal{R}\le(\ell_\kappa/\xi)^2\simeq6$: a large but finite, state-discriminating
signal, not an infinite wall. The absolute depletion scales as $|\Delta\mathcal{E}|\propto\kappa^2$: a
steeper condensate profile (larger $\kappa$) raises it, though it lowers $\ell_\kappa$ and hence
$\mathcal{R}$.

\begin{figure}[t]
 \includegraphics[width=\columnwidth]{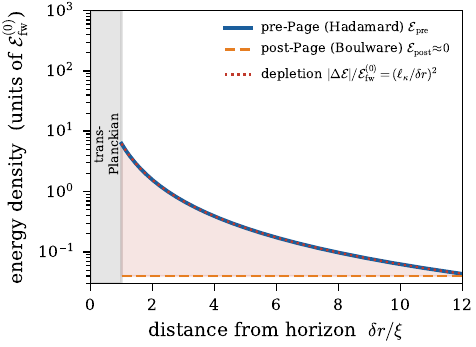}
 \caption{\label{fig:diverge}Near-horizon energy density (units of $\mathcal{E}_{\rm fw}^{(0)}$).
 Pre-Page (Hadamard) the phonon atmosphere is Tolman-blueshifted, $\mathcal{E}_{\rm pre}\propto
 (\delta r)^{-2}$ (blue); post-Page the Boulware-like state is depleted, $\mathcal{E}_{\rm post}\!\approx\!0$
 (orange). The acoustic firewall is this depletion $|\Delta\mathcal{E}|/\mathcal{E}_{\rm fw}^{(0)}=
 (\ell_\kappa/\delta r)^2$ (red), measured differentially across $t_{\rm Page}$. Grey: $\delta r<\xi$,
 where the acoustic metric fails.}
\end{figure}

\begin{figure}[t]
 \includegraphics[width=\columnwidth]{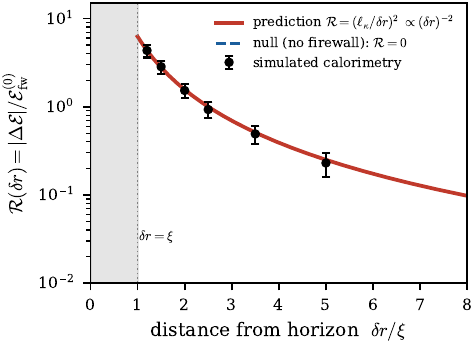}
 \caption{\label{fig:signature}Proposed signature. Predicted depletion ratio
 $\mathcal{R}(\delta r)=(\ell_\kappa/\delta r)^2\propto(\delta r)^{-2}$ (red) versus the no-firewall
 null $\mathcal{R}=0$ (dashed). Points with error bars: simulated differential phonon calorimetry for
 $^{87}$Rb (fixed-seed synthetic data, not measurements). Grey: healing-length cutoff.}
\end{figure}

\emph{Falsifiable prediction.}---The signal is the dimensionless, platform-independent depletion ratio
\begin{equation}
 \mathcal{R}(\delta r)=\frac{|\Delta\mathcal{E}(\delta r)|}{\mathcal{E}_{\rm fw}^{(0)}}
 \ \xrightarrow{\ \delta r\to0\ }\ \left(\frac{\ell_\kappa}{\delta r}\right)^{2},
 \label{eq:R}
\end{equation}
obtained by recording the near-horizon phonon energy-density profile before ($t<t_{\rm Page}$) and after
($t>t_{\rm Page}$) the analogue Page time; the difference isolates the post-Page depletion from static,
state-independent backgrounds. An enhancement rising toward the cutoff as $\mathcal{R}\propto(\delta
r)^{-2}$ over the accessible window $\xi\lesssim\delta r\lesssim\ell_\kappa$, emerging only after
$t_{\rm Page}$, would be evidence for the acoustic firewall [Fig.~\ref{fig:signature}]; a stress tensor
that stays Hadamard across $t_{\rm Page}$ would refute it.
For a $^{87}$Rb condensate our reproducible estimates give $\xi\simeq1~\mu{\rm m}$,
$T_{\rm ac}\simeq1.2\times10^{-10}~{\rm K}$, $\kappa\sim10^{2}~{\rm s^{-1}}$, $\ell_\kappa\simeq2.5\,\xi$,
and an analogue Page time $t_{\rm Page}\sim(R_h/\xi)\,\hbar/(k_BT_{\rm ac})\sim1~{\rm s}$ for a $20~\mu$m
horizon~\cite{AcousticFirewallsCode2026}. The signal is faint---$\mathcal{E}_{\rm fw}^{(0)}$ lies far below
present single-shot calorimetric noise---so many-run averaging is required; a warmer platform such as a
nano-electromechanical horizon ($T_{\rm ac}\sim10^{-3}$~K, $\mathcal{E}_{\rm fw}^{(0)}\sim10^{-14}~{\rm
J/m^3}$) would ease detection, though an analogue horizon with the requisite partner entanglement has yet
to be realized there. The relevant scales otherwise coincide with those of the already-observed analogue
Hawking signal~\cite{Steinhauer2016,deNova2019,Kolobov2021}.

\emph{Discussion.}---The correspondence is exact at the kinematic level---thermal spectrum, partner
entanglement, monogamy---and this is precisely its limit. The acoustic metric~\eqref{eq:metric} is a
fixed background; there are no acoustic Einstein equations, and the quantum backreaction of phonons on
the condensate flow---real and calculable in its own
right~\cite{Schuetzhold2005,BaakRibeiroFischer2022}---does not shrink the horizon, so there is no true
analogue evaporation or literal Page time. What plays the role of $t_{\rm Page}$ is
the time at which the accumulated exterior entropy equals the effective entropy budget of the finite
condensate, after which unitarity of the closed laboratory system enforces Eq.~\eqref{eq:disent}. A
skeptic may object that this is merely Unruh's sonic horizon dressed in AMPS language. We disagree on
one decisive point: the AMPS post-Page mechanism \emph{selects} the Boulware-like vacuum at a laboratory
horizon, turning the otherwise static Boulware/Hartle--Hawking ambiguity into a \emph{time-dependent}
prediction---the post-Page depletion $\mathcal{R}\propto(\delta r)^{-2}$ in Eq.~\eqref{eq:R}, which
neither Unruh's kinematics nor the AMPS argument alone yields and which is absent in every purely thermal
(Hadamard) treatment. The prediction is falsifiable: observing analogue Hawking radiation with a persistently
Hadamard, finite near-horizon stress tensor across $t_{\rm Page}$ would refute it.

Beyond the BEC, the same monogamy obstruction should appear wherever an analogue horizon supports an
entangled partner structure with a finite entropy budget: nano-electromechanical systems, where a
higher $T_{\rm ac}$ and shorter $t_{\rm Page}$ ease detection; slow-light photonic-crystal horizons,
read out interferometrically; and, classically, supersonic water channels~\cite{Weinfurtner2011}. A
positive detection would motivate seeking the analogue island---a turnover of the measured radiation
entropy at $t_{\rm Page}$, the entropic counterpart of the signature reported
here~\cite{AEMM2019,Penington2020,AlmheiriRMP2021}. A caveat: the signal peaks near $\delta r\sim\xi$,
where Bogoliubov dispersion becomes important; whether it sharpens or regulates the depletion is left to
future work~\cite{Balbinot2008,Carusotto2008}.

\emph{Conclusion.}---The monogamy of entanglement, applied to analogue Hawking radiation, selects a
specific non-Hadamard near-horizon phonon state---the acoustic firewall---whose near-horizon thermal
atmosphere, relative to the smooth state, is depleted with a profile growing as $(\delta r)^{-2}$ after
the analogue Page time, giving a concrete, falsifiable differential calorimetric signature
$\mathcal{R}\propto(\delta r)^{-2}$. The
construction imports a central tension of quantum gravity into a controlled laboratory setting; it does
not resolve the information paradox, but it makes one of its sharpest consequences experimentally
addressable; supporting derivations are given in the Supplemental Material.

\begin{acknowledgments}
This work was partially supported by the State Assignment of the Ministry of Education
and Science of the Russian Federation (Project No.\ FZEN 2023-0006), the Nantong Science and Technology
Plan Project (Grant Nos.\ JC2020137 and JC2020138), the Key Research and Development Program of Jiangsu
Province of China (Grant No.\ BE2021013-1), the National Natural Science Foundation of Jiangsu Province
of China (Grant No.\ BK20201438), and in part by the Natural Science Research Project of Jiangsu
Provincial Institutions of Higher Education (Grant Nos.\ 1120KJA510002 and 20KJB510010).
\end{acknowledgments}

\paragraph*{Author contributions.}---Conceptualization, N.S.A.; methodology, N.S.A.; investigation,
S.N.A. and Q.-H.Q.; validation, A.P.N. and Q.-H.Q.; writing---original draft, N.S.A. and A.P.N.;
writing---review and editing, N.S.A. and A.P.N.; supervision, S.N.A.

\paragraph*{Data availability.}---All estimates, symbolic verifications, and figures reported here are
openly available and reproducible from the accompanying repository~\cite{AcousticFirewallsCode2026}
(Zenodo, DOI 10.5281/zenodo.21269544).

\bibliography{refs}

\clearpage
\onecolumngrid
\begin{center}
\textbf{\large Supplemental Material}\\[4pt]
\textbf{\large Acoustic Firewalls: Analogue Gravity Perspective on the AMPS Paradox}
\end{center}
\vspace{6pt}
\twocolumngrid
\setcounter{section}{0}
\setcounter{equation}{0}
\setcounter{figure}{0}
\setcounter{table}{0}
\renewcommand{\thesection}{S\arabic{section}}
\renewcommand{\theequation}{S\arabic{equation}}
\renewcommand{\thefigure}{S\arabic{figure}}
\renewcommand{\thetable}{S\arabic{table}}

This Supplemental Material collects the derivations underlying the Letter. Equation and figure numbers
prefixed by ``S'' are internal; unprefixed numbers refer to the main text. All numerical values and
figures are reproduced by the accompanying repository~\cite{AcousticFirewallsCode2026}.

\section{Acoustic metric and surface gravity}
A barotropic, inviscid, irrotational flow with density $\rho$, velocity $\bm v=\nabla\Phi$, and local
sound speed $c_s$ carries linearized velocity-potential perturbations $\hat\phi$ obeying
$\Box_{\rm ac}\hat\phi=0$, with $\Box_{\rm ac}$ the d'Alembertian of the acoustic metric
[Eq.~(1)]~\cite{Unruh1981,BLV2005}. In a dilute Bose--Einstein condensate (BEC) the Gross--Pitaevskii
equation linearized about a stationary background $\psi_0=\sqrt{n_0}\,e^{i\theta}$ reproduces this
structure, with $\bm v=(\hbar/m)\nabla\theta$, $c_s=\sqrt{g n_0/m}$ ($g=4\pi\hbar^2 a_s/m$), and $\hat\phi$ the phase
fluctuation~\cite{Garay2000,Recati2009}. The horizon is the surface $v(r_h)=c_s$. Linearizing
$c_s^2-v^2$ about $r_h$,
\begin{equation}
 c_s^2-v^2\simeq 2c_s\,\kappa_0\,(r-r_h),\qquad
 \kappa_0=\left.\frac{d(c_s-v)}{dr}\right|_{r_h},
 \label{eq:gtt}
\end{equation}
so the metric coefficient $g_{tt}\propto(c_s^2-v^2)$ vanishes linearly at $r_h$. In the analogue-gravity
rate convention the surface gravity of Eq.~(2) is
$\kappa=(2c_s)^{-1}\partial_r(c_s^2-v^2)|_{r_h}=\kappa_0$ (units $\mathrm{s^{-1}}$); equivalently
$\kappa=(\rho_0 c_s)^{-1}\partial_r P|_{r_h}$ with $P=\tfrac12 g n_0^2$ the condensate pressure. The
surface-gravity length $\ell_\kappa\equiv c_s/(2\kappa)$ enters the near-horizon estimate below.

\section{Analogue Hawking temperature}
Introducing the tortoise coordinate $dr_*/dr=c_s/(c_s^2-v^2)$ and null coordinates
$u=t-r_*/c_s$, $w=t+r_*/c_s$, the near-horizon acoustic line element is conformal to $(1+1)$-dimensional
Minkowski space. A standard Bogoliubov/Unruh--DeWitt-detector calculation then yields a thermal phonon
spectrum with the analogue Hawking temperature of Eq.~(2),
$T_{\rm ac}=\hbar\kappa/(2\pi k_B)$, the acoustic counterpart of $T_H=\hbar\kappa_{\rm grav}/2\pi k_B$.
The correspondence is purely kinematic: it holds at the level of null geodesics and does not invoke the
Einstein equations. The healing length $\xi=\hbar/(\sqrt2\,m c_s)$ sets the scale at which
Bogoliubov dispersion becomes superluminal; low-frequency phonons ($\omega\sim k_B T_{\rm ac}/\hbar\ll
c_s/\xi$) responsible for the thermal signal are insensitive to this modification, as confirmed by
numerics and experiment~\cite{Carusotto2008,deNova2019,Kolobov2021}.

\section{Entangled partner state and thermal reduction}
The Bogoliubov transformation relating in- and out-operators, $\hat a^{\rm out}_\omega=\cosh r_\omega\,
\hat a^{\rm in}_\omega+\sinh r_\omega\,\hat b^{\dagger\,\rm in}_\omega$ with
$\tanh r_\omega=e^{-\hbar\omega/2k_B T_{\rm ac}}$, maps the in-vacuum to the two-mode squeezed state of
Eq.~(3). Tracing over the interior partner $b$ gives the reduced exterior state
\begin{equation}
 \hat\rho_{\rm out}=(1-q)\sum_{n}q^{\,n}\,|n\rangle\langle n|,\qquad q=e^{-\hbar\omega/k_B T_{\rm ac}},
\end{equation}
i.e.\ a thermal (Planck) distribution with $\langle n\rangle=q/(1-q)=[e^{\hbar\omega/k_B T_{\rm ac}}-1]^{-1}$,
verified symbolically in the repository. Its von Neumann entropy
$S(\omega)=-\mathrm{Tr}\,\hat\rho_{\rm out}\ln\hat\rho_{\rm out}$ grows monotonically as phonons are
emitted, the analogue of Bekenstein--Hawking entropy growth. The partner correlations are directly
observable as nonlocal density--density correlations~\cite{Balbinot2008,Steinhauer2016}.

\section{Purity identity and the analogue Page time}
Let $A$ be a late outgoing mode, $B$ its interior partner, and $R$ the early exterior radiation. When
$ABR$ is pure, $S(AR)=S(B)$ and $S(AB)=S(R)$, so the mutual informations obey the exact identity
\begin{equation}
 I(A{:}R)+I(A{:}B)= 2S(A),
 \label{eq:mono}
\end{equation}
the entropic statement of monogamy used by AMPS~\cite{AMPS2013} (the Coffman--Kundu--Wootters and
Osborne--Verstraete inequalities~\cite{CKW2000,OsborneVerstraete2006} are its concurrence-based
relatives). Before the analogue Page time $A$ is maximally entangled with $B$ (smooth Hadamard vacuum),
$I(A{:}B)\to 2S(A)$. Unitarity of the \emph{closed laboratory system}---a condensate of finite atom
number, hence finite entropy budget $S_{\max}\sim(R_h/\xi)^{d-1}$---requires that once the accumulated
exterior entropy reaches $S_{\max}$, further emission must \emph{decrease} $S_{\rm rad}$; this defines
the analogue Page time $t_{\rm Page}$ with no need for evaporation or backreaction. For $t>t_{\rm Page}$
purification forces $I(A{:}R)\to 2S(A)$, and Eq.~\eqref{eq:mono} then gives $I(A{:}B)\to 0$ [Eq.~(5)]:
the partner entanglement that guaranteed a Hadamard state at $r_h$ is destroyed. An order-of-magnitude
estimate $t_{\rm Page}\sim(R_h/\xi)\,\hbar/(k_B T_{\rm ac})\sim 1~\mathrm{s}$ for a $20~\mu$m $^{87}$Rb
horizon is computed in the repository.

\section{Renormalized stress tensor: Hadamard vs.\ non-Hadamard state}
This is the central technical point. In the $(1+1)$-dimensional near-horizon theory, the renormalized
flux of a conformal scalar in a state $|\Omega\rangle$ is obtained by point splitting,
\begin{equation}
 \langle\hat T_{uu}\rangle_\Omega=\lim_{u'\to u}\big[-\partial_u\partial_{u'}
 G^{(1)}_\Omega+\text{(Hadamard)}\big].
\end{equation}
For the Hartle--Hawking--Unruh (Hadamard) state the coincidence-limit $(u-u')^{-2}$ singularity of
$G^{(1)}$ is state-independent and exactly removed by the subtraction, leaving the finite thermal flux
equal to the $(1+1)$D Hawking luminosity~\cite{ChristensenFulling1977,BirrellDavies1982}
\begin{equation}
 \langle\hat T_{uu}\rangle_{\rm H}=L_H=\frac{\pi k_B^2 T_{\rm ac}^2}{12\hbar}
 =\frac{\hbar\kappa^2}{48\pi},
 \label{eq:LH}
\end{equation}
verified symbolically in the repository. Severing the partner entanglement [Eq.~(5)] selects instead a
\emph{Boulware-like}, non-Hadamard state (no outgoing partner correlations across $r_h$). Working in the
near-horizon $(1+1)$D theory conformal to $ds^2=-f\,du\,dv$ with $f=(c_s^2-v^2)/c_s^2$, the trace-anomaly
method fixes the renormalized components up to state-dependent chiral functions $t_u(u),t_v(v)$
\cite{ChristensenFulling1977,BirrellDavies1982,FabbriNavarroSalas2005}. The Boulware-like state
($t_u=t_v=0$, empty as $\delta r\to\infty$) and the Hartle--Hawking state give
\begin{equation}
 \langle\hat T_{uu}\rangle_{\rm B}=-\frac{\hbar\,(f'^2-2ff'')}{192\pi},\qquad
 \langle\hat T_{uu}\rangle_{\rm H}=\langle\hat T_{uu}\rangle_{\rm B}+\frac{\hbar\kappa^2}{48\pi},
 \label{eq:Tuu2d}
\end{equation}
the constant $\hbar\kappa^2/48\pi=-\tfrac{1}{24\pi}\{U,u\}$ being minus the Schwarzian of the Kruskal map
$U=-\kappa^{-1}e^{-\kappa u}$. At $r_h$ ($f\to0$, $f'\to2\kappa$) one has $\langle\hat T_{uu}\rangle_{\rm H}\to0$
(regular, Hadamard) while $\langle\hat T_{uu}\rangle_{\rm B}\to-\hbar\kappa^2/48\pi$; as $\delta r\to\infty$,
$\langle\hat T_{uu}\rangle_{\rm B}\to0$ and $\langle\hat T_{uu}\rangle_{\rm H}\to\hbar\kappa^2/48\pi=L_H$,
recovering Eq.~\eqref{eq:LH}. The disentangled-minus-Hadamard difference is therefore the \emph{constant,
negative} null flux
\begin{equation}
 \langle\hat T_{uu}\rangle_{\rm B}-\langle\hat T_{uu}\rangle_{\rm H}=-\frac{\hbar\kappa^2}{48\pi}<0,
 \label{eq:diff2d}
\end{equation}
a genuine depletion whose coefficient is fixed \emph{from first principles} (verified symbolically in the
repository). The renormalized null components are thus finite in static $(t,r)$ coordinates; projected on a
static orthonormal frame the difference Tolman-enhances as $\hbar\kappa^2/(24\pi f)\propto(\delta r)^{-1}$
($\propto T_{\rm loc}^2$, the $(1+1)$D result), and in an infalling frame it blows up faster still---the
textbook pathology of the Boulware vacuum~\cite{BirrellDavies1982,FabbriNavarroSalas2005}.

Physically, the disentanglement \emph{removes} the near-horizon thermal atmosphere: the Hadamard
(pre-Page) state carries the Tolman-blueshifted phonon atmosphere $\mathcal{E}_{\rm pre}$, while the
Boulware-like (post-Page) state is nearly empty, $\mathcal{E}_{\rm post}\approx0$. The observable is the
post-Page change $\Delta\mathcal{E}=\mathcal{E}_{\rm post}-\mathcal{E}_{\rm pre}\approx-\mathcal{E}_{\rm pre}$,
isolated by differential calorimetry across $t_{\rm Page}$, which cancels any static, state-independent
background (this is what the subtraction achieves---not a cancellation between two coincident profiles).
The local temperature seen at $\delta r=r-r_h$ is, using Eq.~\eqref{eq:gtt},
\begin{equation}
 T_{\rm loc}(\delta r)=\frac{T_{\rm ac}\,c_s}{\sqrt{c_s^2-v^2}}
 \;\xrightarrow[\delta r\to0]{}\; \frac{T_{\rm ac}}{\sqrt{2\kappa\,\delta r/c_s}},
\end{equation}
so the pre-Page three-dimensional thermal density $\mathcal{E}_{\rm pre}\sim(k_B T_{\rm loc})^4/(\hbar^3
c_s^3)$---and hence the near-complete depletion $|\Delta\mathcal{E}|\simeq\mathcal{E}_{\rm pre}$---is
\begin{equation}
 \frac{|\Delta\mathcal{E}(\delta r)|}{\mathcal{E}_{\rm fw}^{(0)}}
 =\left(\frac{c_s^2}{c_s^2-v^2}\right)^{\!2}
 \;\xrightarrow[\delta r\to0]{}\;\left(\frac{\ell_\kappa}{\delta r}\right)^{\!2},
 \qquad \ell_\kappa=\frac{c_s}{2\kappa},
 \label{eq:tolman}
\end{equation}
a \emph{negative} contribution (a depletion of the near-horizon atmosphere), regulated where the
linearization $c_s^2-v^2\simeq2c_s\kappa\,\delta r$ ceases to apply, i.e.\ at $\delta r\sim\xi$. Here
$\delta r=r-r_h$ is the radial \emph{coordinate} distance and $\mathcal{E}$ is the proper energy density
read by a static three-dimensional calorimeter. The exponent $-2$ is fixed by the three-dimensionality of
the phonon gas and the linear vanishing of $g_{tt}$; the prefactor scale is $\ell_\kappa$, not $\xi$. We
emphasize the status of this result: the $(1+1)$D coefficient [Eq.~\eqref{eq:diff2d}] and the Tolman law
$T_{\rm loc}=T_{\rm ac}/\sqrt{f}$ are rigorous, whereas the transverse-mode counting that promotes the
static-frame exponent from $-1$ [the $(1+1)$D value] to $-2$ [the $3$D thermal gas
$\mathcal{E}\sim(k_BT_{\rm loc})^4/(\hbar^3c_s^3)$] is a Stefan--Boltzmann thermodynamic extension,
consistent with the known near-horizon behavior of the Boulware stress tensor in $3{+}1$
dimensions~\cite{Candelas1980}; we do not claim a closed-form $3{+}1$D renormalized
$\langle\hat T_{\mu\nu}\rangle$ on the acoustic background. In the regime of validity
$\delta r\gtrsim\xi$ the enhancement is bounded, $|\Delta\mathcal{E}|/\mathcal{E}_{\rm fw}^{(0)}\le
(\ell_\kappa/\xi)^2\simeq6$: the growth is formal (a state-discriminator), not an infinite wall. The
dimensionless depletion ratio $\mathcal{R}(\delta r)=|\Delta\mathcal{E}|/\mathcal{E}_{\rm fw}^{(0)}$
[Eq.~(9)] is platform independent and is the proposed observable. Whether Bogoliubov dispersion at
$\delta r\sim\xi$ sharpens or further regulates the depletion is a quantitative question for the
nonlocal density--density correlator~\cite{Balbinot2008,Carusotto2008}.

\section{Parametric estimates and platforms}
\begin{figure}[t]
 \includegraphics[width=0.9\columnwidth]{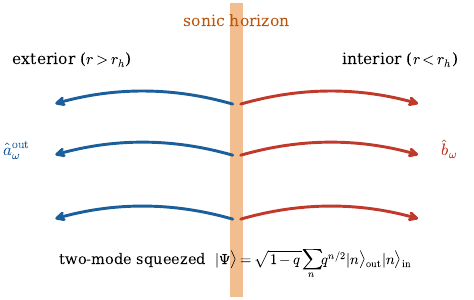}
 \caption{\label{fig:pairs}Phonon pair production at the sonic horizon: each outgoing mode
 $\hat a^{\rm out}_\omega$ is entangled with its interior partner $\hat b_\omega$ in the two-mode
 squeezed state of Eq.~(3).}
\end{figure}

\begin{figure}[t]
 \includegraphics[width=\columnwidth]{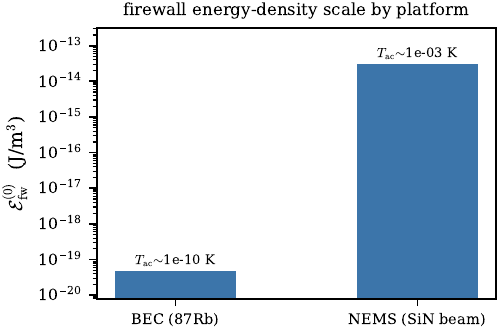}
 \caption{\label{fig:platforms}Firewall energy-density scale $\mathcal{E}_{\rm fw}^{(0)}$ across
 platforms with a well-defined analogue temperature. The higher $T_{\rm ac}$ of a NEMS horizon yields a
 far larger $\mathcal{E}_{\rm fw}^{(0)}$ and a shorter analogue Page time, at the cost of a colder
 partner-entanglement readout.}
\end{figure}

For $^{87}$Rb ($m=87\,\mathrm{u}$, $a_s\simeq100\,a_0$, $n_0\sim10^{13}\,\mathrm{cm^{-3}}$,
$c_s\simeq0.5\,\mathrm{mm/s}$) the repository gives $\xi\simeq1.0\,\mu\mathrm{m}$,
$\kappa\sim10^{2}\,\mathrm{s^{-1}}$, $T_{\rm ac}\simeq1.2\times10^{-10}\,\mathrm{K}$,
$\mathcal{E}_{\rm fw}^{(0)}\simeq5\times10^{-20}\,\mathrm{J/m^3}$, and $t_{\rm Page}\sim1\,\mathrm{s}$.
Table~\ref{tab:platforms} compares candidate platforms; Figs.~\ref{fig:pairs} and \ref{fig:platforms}
illustrate the partner structure and the platform hierarchy. Beyond these bosonic systems, fermionic
Weyl-semimetal analogues---where a horizon emerges when the Weyl cone tilts into the type-II
regime---realize black- and white-hole horizons with far higher effective Hawking
temperatures~\cite{Volovik2016,Kedem2020,Yang2025}; the present observable, however, relies on bosonic
phonon calorimetry and does not carry over directly to a fermionic spectrum.

\begin{table*}[t]
\caption{\label{tab:platforms}Candidate platforms for the acoustic firewall signature. Speeds are the
relevant $c_s$ or group velocity $v_g$; $\mathcal{E}_{\rm fw}^{(0)}$ is the energy-density scale
[Eq.~(6)]. A dash (``---'') marks a quantity that is not defined for that platform; the reason is given
in the corresponding footnote. NEMS: nano-electromechanical system; SNSPD: superconducting-nanowire
single-photon detector.}
\begin{ruledtabular}
\begin{tabular}{lcccl}
Platform & speed & $T_{\rm ac}$ & $\mathcal{E}_{\rm fw}^{(0)}$ & Detector \\
\hline
BEC ($^{87}$Rb) & $0.5$\,mm/s & $10^{-10}$\,K & $5\times10^{-20}$\,J/m$^3$ & phonon calorimetry \\
NEMS (SiN) & $10^{4}$\,m/s & $10^{-3}$\,K & $3\times10^{-14}$\,J/m$^3$ & microwave optomechanics \\
Photonic crystal & $c/n_g$ & ---\footnote{The photonic-crystal analogue is strongly dispersive: its
effective Hawking temperature is fixed by the specific waveguide design rather than by a single
near-horizon surface gravity $\kappa$, so no representative $T_{\rm ac}$---and hence no
$\mathcal{E}_{\rm fw}^{(0)}$---is quoted for the platform class.} & ---\footnotemark[1] & SNSPD + HOM \\
Water channel & $0.3$\,m/s & classical & ---\footnote{The water channel is a classical (stimulated)
surface-wave analogue, as indicated by the ``classical'' entry in the $T_{\rm ac}$ column; the vacuum
energy-density scale $\mathcal{E}_{\rm fw}^{(0)}\propto\hbar$ has no classical counterpart.} &
surface-wave correlator \\
\end{tabular}
\end{ruledtabular}
\end{table*}

\section{Experimental protocol}
The signature is extracted differentially, cancelling static backgrounds: (i) prepare the analogue
horizon in the in-vacuum; (ii) record the phonon energy-density profile
$\langle\hat T_{uu}\rangle(\delta r)$ at $t<t_{\rm Page}$ (Hadamard baseline); (iii) record it again at
$t>t_{\rm Page}$; (iv) form the ratio $\mathcal{R}(\delta r)$ of Eq.~(9). A power law
$\mathcal{R}\propto(\delta r)^{-2}$ emerging only in the post-Page data, and cut off at $\delta r\sim\xi$,
is the acoustic firewall. A complementary test monitors the radiation entropy $S_{\rm rad}(t)$ by phonon
tomography: a turnover at $t_{\rm Page}$ (the analogue Page curve) is the entropic counterpart of the
stress-tensor signature and would indicate an analogue ``island.''

\typeout{get arXiv to do 4 passes: Label(s) may have changed. Rerun}
\end{document}